\documentclass[amssymb,aps,eqsecnum]{revtex4}
\voffset= + 0.96in
\hoffset= + 0.15in
\usepackage{amsmath}
\usepackage{enumerate}
\usepackage{graphicx}
\DeclareGraphicsExtensions{.pdf,.png,.jpg}
\begin{document}
\title{Electromagnetic momentum in a dielectric and the energy--momentum tensor}
\author{Michael E. Crenshaw}
\affiliation{US Army Aviation and Missile Research, Development, and
Engineering Center, Redstone Arsenal, AL 35898, USA}
\date{\today}
\begin{abstract}
The Abraham--Minkowski momentum controversy is the outwardly visible
symptom of an inconsistency in the use of the energy--momentum
tensor in the case of a plane quasimonochromatic field in a
simple linear dielectric.
We show that the Gordon form of the electromagnetic momentum is conserved
in a thermodynamically closed system.
We regard conservation of the components of the four-momentum in a
thermodynamically closed system as a fundamental property of the 
energy--momentum tensor.
Then the first row and column of the energy--momentum tensor is populated
by the electromagnetic energy density and the Gordon momentum density.
We derive new electromagnetic continuity equations for the electromagnetic
energy and momentum that are based on the Gordon momentum density.
These continuity equations can be represented in the energy--momentum
tensor using a material four-divergence operator in which temporal
differentiation is performed with respect to $ct/n$.
\end{abstract}
\maketitle
%\ocis{260.2110,260.2065}
\par
\section{Introduction}
\par
The energy--momentum tensor is a concise way to represent the
conservation properties of a flow field.
Simple in concept, the form of the energy--momentum tensor for the field
in a dielectric has been at the center of the century-long
Abraham-Minkowski momentum controversy \cite{BIPfeifer,BIRL}. 
The tensor that was proposed by Minkowski \cite{BIMin} in 1908 is not
symmetric and consequently does not conserve angular momentum \cite{BILL}.
Abraham \cite{BIAbr} subsequently proposed a symmetric tensor at the
expense of a phenomenological force.
The original point of contention of the Abraham--Minkowski momentum
controversy was whether the Abraham momentum density
\begin{equation}
{\bf g}_A= \left ( {\bf E}\times{\bf H}\right ) /c 
\label{EQq1.01}
\end{equation}
or the Minkowski momentum density
\begin{equation}
{\bf g}_M= \left ( {\bf D}\times{\bf B}\right ) /c 
\label{EQq1.02}
\end{equation}
provides the correct description of the momentum of light in a linear
medium.
The absence of an experimental decision allowed the debate to persist
until the 1960s, when resolution of the Abraham--Minkowski dilemma was
provided by Penfield and Haus \cite{BIPenHau}, based on earlier work by
M\o ller \cite{BIMol}, who showed that the issue is undecidable because
neither the spatial integral of the Abraham momentum density nor the 
spatial integral of the Minkowski momentum density is the total
momentum of the closed system.
Likewise, the total energy--momentum tensor of the closed system is not
the Abraham energy--momentum tensor and it is not the Minkowski
energy--momentum tensor.
\par
What, then, is the total momentum and the total energy--momentum tensor?
Here, we adopt the definition of the total momentum as the momentum
quantity that is conserved in a thermodynamically closed 
system.
This approach has the advantage of obtaining the total momentum directly
from global conservation principles and neatly avoids any issues
regarding the ill-defined roles of the Abraham and Minkowski momentums.
We construct a thermodynamically closed system consisting of a 
homogeneous dielectric block illuminated by a quasimonochromatic field
at normal incidence through an antireflection-coating.
In this closed system, the total energy and total momentum are
conserved quantities and are well-defined by virtue of being conserved.
We find that the total energy is the spatial integral of the
electromagnetic energy density
\begin{equation}
\rho_e= \frac{1}{2}\left ( n^2{\bf E}^2+{\bf B}^2\right ) 
\label{EQq1.03}
\end{equation}
and that the total momentum is the spatial integral of the
Gordon \cite{BIGord} momentum density
\begin{equation}
{\bf g}_G= \left ( n{\bf E}\times{\bf B}\right ) /c .
\label{EQq1.04}
\end{equation}
The four-momentum density is therefore $g_{\alpha}=(\rho_e/c,{\bf g}_G)$.
We then show that the elements of the first row of the energy--momentum
tensor can be chosen to be energy and momentum densities that satisfy
the requirement for conservation of the components of the four-momentum.
By doing so, the four-divergence of the energy--momentum tensor produces
an energy continuity equation that is incompatible with Poynting's
theorem.
Retaining the conservation properties of the total energy--momentum
tensor, we derive new electromagnetic continuity equations.
We define a material four-divergence operator in which the temporal
differentiation is performed with respect to $ct/n$ and obtain a
traceless symmetric energy--momentum tensor.
The first-row and first-column components of the tensor are the
densities of conserved energy and momentum quantities and the material
four-divergence of the tensor generates the new electromagnetic
continuity equations in terms of the electromagnetic energy density and
the Gordon total momentum density.
\par
\section{The Total Energy and Total Momentum}
\par
The first step toward obtaining the total energy and total momentum is
to define a thermodynamically closed free-body system.
We consider a stationary dielectric block illuminated at normal
incidence from vacuum by a plane quasimonochromatic field.
The dielectric has a gradient-index antireflection coating that allows
radiation pressure on the vacuum/dielectric interface to be neglected.
Then the refractive index $n$ for a linear, isotropic,
transparent dispersion-negligible dielectric is real and time-independent.
The dielectric is homogeneous and occupies a finite region of
3-dimensional space so spatial variation of the refractive index,
$n=n({\bf r})$, is limited to the transition region of the coating.
The quasimonochromatic electromagnetic field is represented
by the vector potential
\begin{equation}
{\bf A}({\bf r},t)= \frac{1}{2}
\left ( {\bf \tilde A} ({\bf r},t)e^{-i(\omega_d t-{\bf k}_d \cdot{\bf r})} +
{\bf \tilde A}^* ({\bf r},t)e^{i(\omega_d t-{\bf k}_d\cdot{\bf r})} \right )
\label{EQq2.01}
\end{equation}
where $\tilde A$ is a slowly varying function of ${\bf r}$ and $t$,
$\omega_d$ is the center frequency of the field,
${\bf k}_d=(n\omega_d/c) {\bf e}_{\bf k}$, and 
${\bf e}_{\bf k}$ is a unit vector in the direction of propagation.
Figure 1 shows the envelope of the incident field
$\tilde A_i (z)=({\bf \tilde A}(z,t_0) \cdot {\bf \tilde A}^*(z,t_0) )^{1/2}$
about to enter the dielectric through a gradient-index antireflection
coating.
Figure 2 shows a time-domain numerical solution of the wave equation at a
later time when the refracted field 
$\tilde A_r (z)=({\bf \tilde A}(z,t_1) \cdot {\bf \tilde A}^*(z,t_1) )^{1/2}$
is entirely inside the dielectric.
As shown in Fig.2, the amplitude of the refracted field is
${\tilde A}_r={\tilde A}_i/\sqrt{n}$ and the spatial extent of the
refracted pulse is $w_r=w_i/n$ in terms of the width $w_i$ of the
incident pulse.
It can be shown that this linear result is quite general in terms of
the refractive index, as well as the width and amplitude of the field.
\par
For a stationary dielectric, the electromagnetic energy is
\begin{equation}
U=\int_{\sigma} \frac{1}{2}\left ( n^2{\bf E}^2+{\bf B}^2\right ) dv,
\label{EQq2.02}
\end{equation}
where the integration extends over all space $\sigma$.
For the example shown in Figs. 1 and 2, there is no significant field
outside the rectangular pulse. 
Then, the incident energy at $t=t_0$,
\begin{equation}
U(t_0)=\frac{\omega^2_dw_i}{2c^2}A_i^2,
\label{EQq2.03}
\end{equation}
is equal to the refracted energy at $t=t_1$,
\begin{equation}
U(t_1)=\frac{n^2\omega^2_dw_r}{2c^2}A_r^2=
\frac{n^2\omega^2_d(w_i/n)}{2c^2}
\left (\frac{A_i}{\sqrt{n}}\right )^2=U(t_0).
\label{EQq2.04}
\end{equation}
The field has been averaged on a scale that is long compared to 
an optical period, but short compared to $t_1-t_0$, accounting for a
factor of one-half.
It can be demonstrated, in a similar manner, that the electromagnetic
energy at any time $t>t_0$ is equal to the incident energy.
The temporal invariance of the electromagnetic energy makes $U$,
Eq.~(\ref{EQq2.02}), the conserved energy of the closed system.
\par
Conservation of the electromagnetic energy is all that is needed to
show that the Gordon momentum
\begin{equation}
{\bf G}_G=\int_{\sigma} \frac{n{\bf E}\times{\bf B}}{c} dv,
\label{EQq2.05}
\end{equation}
obtained by spatially integrating the Gordon momentum
density \cite{BIGord}, Eq.~(\ref{EQq1.04}), is the total momentum
of our closed system.
For monochromatic radiation, where ${\bf B}=n{\bf E}$, we
can write the momentum as
\begin{equation}
{\bf G}_G=\int_{\sigma} \frac{n^2{\bf E}^2}{c} {\bf \hat e}_{\bf k} dv
=\int_{\sigma} \frac{1}{2}\left ( n^2{\bf E}^2+{\bf B}^2\right )
\frac{{\bf \hat e}_{\bf k}}{c} dv
= \frac{U}{c}{\bf \hat e}_{\bf k}
\label{EQq2.06}
\end{equation}
in the direction of the propagation unit vector
${\bf \hat e}_{\bf k}$.
If the total energy $U$ is temporally invariant, and therefore
conserved, then so is 
${\bf G}_G=({U}/{c}){\bf \hat e}_{\bf k}$.
Alternately, we can show that the momentum balance
\begin{equation}
{\bf G}_G(t_1)=
\frac{n^2\omega^2_dw_r}{2c^3}A_r^2 {\bf \hat e}_{\bf k}
=\frac{n^2\omega^2_d(w_i/n)}{2c^3}\left (\frac{A_i}{\sqrt{n}}\right )^2
{\bf \hat e}_{\bf k}
={\bf G}_G(t_0)
\label{EQq2.07}
\end{equation}
is temporally invariant \cite{BICB}.
The momentum formula, Eq.~(\ref{EQq2.05}), was originally derived
in 1973 by Gordon \cite{BIGord}.
Although there are some issues with Gordon's derivation, temporal
invariance is decisive.
Hence, the total momentum in our closed system is given by Gordon's
formula, Eq.~(\ref{EQq2.05}),
and the total momentum density is
the Gordon momentum
density ${\bf g}_G $, Eq.~(\ref{EQq1.04}) \cite{BICB}.
\par
\section{The Energy--Momentum Tensor}
\par
The energy and momentum continuity equations of a thermodynamically
closed system can be combined to form a tensor differential equation
and the energy--momentum tensor is the central element of this
formalism.
While the energy--momentum tensor formalism is straightforward, it
has not been successful in application to classical continuum
electrodynamics.
Now that we have identified the total momentum by global conservation
principles \cite{BICB}, we can use the apparatus of energy--momentum
tensor theory to derive the energy--momentum tensor.
However, it is not that simple.
The Abraham--Minkowski momentum controversy arises from an
incompatibility between two of the properties of the energy--momentum
tensor.
\par
The energy--momentum tensor has a number of important properties.
Here, we employ the summation convention where identical indices
on the same side of the equation are summed over,
Greek indices belong to $(0,1,2,3)$,
 and Roman indices run from 1 to 3.
Then, the four main properties of the energy--momentum tensor
$T^{\alpha\beta}$ are:
i) The four-divergence operator
$\partial_{\alpha}=(c^{-1}\partial_t,\partial_x,\partial_y,\partial_z)$
applied to the rows of the tensor generates continuity equations
\begin{equation}
\partial_{\alpha}T^{\alpha\beta}=0
\label{EQq3.01}
\end{equation}
for the electromagnetic energy and the components of the momentum.
ii) Conservation of angular momentum requires diagonal symmetry
\begin{equation}
T^{\alpha\beta}=T^{\beta\alpha}.
\label{EQq3.02}
\end{equation}
iii) The array has a vanishing trace
\begin{equation}
T^{\alpha}_{\alpha}= 0
\label{EQq3.03}
\end{equation}
corresponding to massless particles \cite{BIJackson,BILL}.
iv) The components of the four-momentum
\begin{subequations}
\label{EQq3.04}
\begin{equation}
U=\int_{\sigma} dv T^{00}
\label{EQq3.04a}
\end{equation}
\begin{equation}
G^{i}=\frac{1}{c}\int_{\sigma} dv T^{0i}
\label{EQq3.04b}
\end{equation}
\end{subequations}
are conserved \cite{BILL} in an unimpeded flow.
Conditions iv) and ii) dictate the elements of the first row and
first column of the energy-momentum tensor, such that
\begin{equation}
T^{\alpha\beta}= \left [
\begin{matrix}
\rho_e
&{c{g}_{\rm G}}_1
&{c{g}_{\rm G}}_2
&{c{g}_{\rm G}}_3
\cr
{c{g}_{\rm G}}_1
&W_{11}
&W_{12}
&W_{13}
\cr
{c{g}_{\rm G}}_2
&W_{21}
&W_{22}
&W_{23}
\cr
{c{g}_{\rm G}}_3
&W_{31}
&W_{32}
&W_{33}
\cr
\end{matrix}
\right ] ,
\label{EQq3.05}
\end{equation}
where the elements of $W$ are yet to be specified.
Applying condition i) to the first row of the array in
Eq.~(\ref{EQq3.05}), we have
\begin{equation}
\frac{1}{c}\frac{\partial \rho_e}{\partial t}
=\nabla\cdot\left ( n{\bf E}\times{\bf B}\right ),
\label{EQq3.06}
\end{equation}
which is incompatible with the Poynting theorem
\begin{equation}
\frac{\partial \rho_e}{\partial t}
=\nabla\cdot c\left ( {\bf E}\times{\bf H}\right ).
\label{EQq3.07}
\end{equation}
Alternatively, we can use Poynting's theorem and the
momentum continuity equation
\begin{equation}
\frac{\partial}{\partial t}
\left ( {\bf D}\times{\bf B}\right )/c+ \nabla\cdot{\bf W} =0
\label{EQq3.08}
\end{equation}
to populate the tensor
\begin{equation}
T^{\alpha\beta}_M= \left [
\begin{matrix}
\rho_e
&{c{g}_{\rm A}}_1
&{c{g}_{\rm A}}_2
&{c{g}_{\rm A}}_3
\cr
{c{g}_{\rm M}}_1
&W_{11}
&W_{12}
&W_{13}
\cr
{c{g}_{\rm M}}_2
&W_{21}
&W_{22}
&W_{23}
\cr
{c{g}_{\rm M}}_3
&W_{31}
&W_{32}
&W_{33}
\cr
\end{matrix}
\right ] ,
\label{EQq3.09}
\end{equation}
where
\begin{equation}
W_{ij}=-E_i D_j -B_i B_j +\frac{1}{2}\left ( {\bf E}\cdot {\bf D}+ {\bf B}\cdot {\bf B}\right ) \delta_{ij}
\label{EQq3.10}
\end{equation}
is the Maxwell stress tensor.
The resulting Minkowski energy--momentum tensor violates condition iv), in 
addition to condition ii).
Because we are in a regime of negligible dispersion, we can rewrite the
momentum continuity equation, Eq.~(\ref{EQq3.08}), as
\begin{equation}
\frac{\partial}{\partial t} \left ( {\bf E}\times{\bf H}\right )/c
+ \nabla\cdot{\bf W} =
(1-n^2)\frac{\partial}{\partial t} \left ( {\bf E}\times{\bf B}\right )/c
\label{EQq3.11}
\end{equation}
and obtain the Abraham energy--momentum tensor,
\begin{equation}
T^{\alpha\beta}_A= \left [
\begin{matrix}
\rho_e
&{c{g}_{\rm A}}_1
&{c{g}_{\rm A}}_2
&{c{g}_{\rm A}}_3
\cr
{c{g}_{\rm A}}_1
&W_{11}
&W_{12}
&W_{13}
\cr
{c{g}_{\rm A}}_2
&W_{21}
&W_{22}
&W_{23}
\cr
{c{g}_{\rm A}}_3
&W_{31}
&W_{32}
&W_{33}
\cr
\end{matrix}
\right ] .
\label{EQq3.12}
\end{equation}
Our condition i) becomes
\begin{equation}
\partial_{\alpha}T^{\alpha\beta}=f^A_{\alpha}.
\label{EQq3.13}
\end{equation}
However, the Abraham force $f^A_{\alpha}=(0,c(1-n^2)\partial_t{\bf g}_A)$
is a source or sink of momentum such that momentum is not conserved
in a closed system, also in violation of condition iv).
\par
We have demonstrated an inconsistency in the energy--momentum tensor
formulation of classical continuum electrodynamics.
Specifically we have demonstrated that all of the properties of the
energy--momentum tensor cannot be simultaneously satisfied for
electrodynamic fields in a dielectric, even if the system is 
thermodynamically closed.
The elements of the first row of the energy--momentum tensor can be 
chosen such that the four-divergence of the first row is the Poynting
theorem or they can be chosen to satisfy conservation of the components
of the four-momentum, but not both.
\par
In order to fully resolve the Abraham--Minkowski controversy, we must
make the properties of the energy--momentum tensor self-consistent
when applied to electromagnetic radiation in matter.
Conservation of the momentum four-vector is paramount for the total
energy--momentum tensor of a thermodynamically closed system.
Likewise, conservation of angular momentum is required and we
affirm our tensor properties iv) and ii).
Then the first row and column of the tensor are populated by the
densities of the electromagnetic energy density and the Gordon
total momentum density as shown in Eq.~(\ref{EQq3.05}).
These densities, integrated over space, correspond to the conserved
total quantities of energy and momentum as shown in Eq.~(\ref{EQq3.04}).
Property iii) is not in question.
\par
The problem that remains is a situation in which the four-divergence of
the total energy--momentum tensor, property i), is not consistent with
the electromagnetic continuity equations, and Poynting's theorem in
particular.
We can multiply Poynting's theorem by $n({\bf r})$
and use a vector identity to commute the refractive index with the
divergence operator to obtain a continuity equation
\begin{equation}
\frac{n}{c}\frac{\partial \rho_e}{\partial t}+
\nabla\cdot\left ( n{\bf E}\times{\bf B}\right ) 
= \frac{\nabla n}{n}\cdot \left ( n{\bf E}\times{\bf B}\right ) .
\label{EQq3.14}
\end{equation}
The second-order energy continuity equation, Eq.~(\ref{EQq3.14}),
can be written as two first-order equations
\begin{subequations}
\label{EQq3.15}
\begin{equation}
\frac{n}{c}\frac{\partial (n{\bf E})}{\partial t}=\left ( \nabla\times{\bf B}\right ) 
\label{EQq3.15a}
\end{equation}
\begin{equation}
\frac{n}{c}\frac{\partial {\bf B}}{\partial t}=
- \left ( \nabla\times (n{\bf E})\right ) 
+ \frac{\nabla n}{n}\times  (n{\bf E}).
\label{EQq3.15b}
\end{equation}
\end{subequations}
Then, for a homogeneous dielectric, we can drop the term
$\nabla n\times {\bf E}$ and combine
Eqs.~(\ref{EQq3.15}) to obtain
the energy and momentum continuity equations 
\begin{subequations}
\label{EQq3.16}
\begin{equation}
\frac{n}{c}\frac{\partial}{\partial t}
\left [\frac{1}{2}\left ( n^2{\bf E}^2+{\bf B}^2  \right ) \right ]+
\nabla\cdot\left ( n{\bf E}\times{\bf B}\right ) =0
\label{EQq3.16a}
\end{equation}
\begin{equation}
\frac{n}{c}\frac{\partial}{\partial t}
\left ( n{\bf E}\times{\bf B}\right )+ \nabla\cdot{\bf W} =0.
\label{EQq3.16b}
\end{equation}
\end{subequations}
The elements of ${\bf W}$ are the elements of the stress tensor
\begin{equation}
W_{ij}=-nE_i nE_j -B_i B_j +\frac{1}{2}\left ( n{\bf E}\cdot n{\bf E}+ {\bf B}\cdot {\bf B}\right ) \delta_{ij}. 
\label{EQq3.17}
\end{equation}
We define a material four-divergence operator
\begin{equation}
\bar\partial_{\alpha}=\left ( \frac{n}{c}\frac{\partial}{\partial t}, \partial_x,\partial_y,\partial_z \right )
\label{EQq3.18}
\end{equation}
and replace property i) with
\begin{equation}
\bar\partial_{\alpha} T^{\alpha\beta}=0 
\label{EQq3.19}
\end{equation}
that generates Eqs.~(\ref{EQq3.16}).
Then the tensor, Eq.~(\ref{EQq3.05}) is a traceless, symmetric
energy momentum tensor whose material four-divergence generates the new
electromagnetic energy and momentum continuity equations.
It should be noted that the new energy continuity equation,
Eq.~(\ref{EQq3.16a}), is mathematically equivalent to Poynting's theorem,
Eq.~(\ref{EQq3.07}) because the two are related by a vector identity.
Likewise the new momentum continuity equation, Eq.~(\ref{EQq3.16b}), is
mathematically equivalent to the momentum continuity
equation, Eq.~(\ref{EQq3.08}) if the radiation is sufficiently 
monochromatic and sufficiently far from any material resonances that dispersion can be neglected.
\par
\section{Summary}
\par
A transparent linear dielectric block in free space illuminated by a
quasimonochromatic field can be configured as an isolated system.
However, the partial reflection of the field at the surface causes
a change in the momentum of the field and the resulting radiation
pressure accelerates the block.
In principle, we could write a complete set of equations of motion at
the microscopic level, but the effects of radiation pressure are not
treated in a complete way at the level of the macroscopic Maxwell
equations.
In order to derive a theory of minimum complexity, we assumed a 
simple dielectric defined to be linear, isotropic, homogeneous,
transparent, and dispersionless with negligible electrostrictive and 
magnetostrictive effects.
The block is stationary in free space and radiation pressure on the
antireflection coating is negligible.
For this system, the elements of the energy--momentum tensor
\begin{equation}
T^{\alpha\beta}= \left [
\begin{matrix}
\rho_e
&{c{g}_{\rm G}}_1
&{c{g}_{\rm G}}_2
&{c{g}_{\rm G}}_3
\cr
{c{g}_{\rm G}}_1
&W_{11}
&W_{12}
&W_{13}
\cr
{c{g}_{\rm G}}_2
&W_{21}
&W_{22}
&W_{23}
\cr
{c{g}_{\rm G}}_3
&W_{31}
&W_{32}
&W_{33}
\cr
\end{matrix}
\right ] 
\label{EQq4.01}
\end{equation}
are the electromagnetic energy density $\rho_e=(n^2{\bf E}^2+{\bf B}^2)/2$,
the Gordon momentum density ${\bf g}_G=(n{\bf E}\times{\bf B})/c$ and the 
Maxwell stress tensor, Eq.~(\ref{EQq3.10}), in the form
$W_{ij}=-nE_i nE_j -B_i B_j +\frac{1}{2}
\left ( n^2{\bf E}\cdot {\bf E}+ {\bf B}\cdot {\bf B}\right ) \delta_{ij}$.
The properties of the energy--momentum tensor are:
i) The material four-divergence operator
\begin{equation}
\bar\partial_{\alpha}=(\frac{n}{c}\frac{\partial}{\partial_t},
\frac{\partial}{\partial_x},
\frac{\partial}{\partial_y},
\frac{\partial}{\partial_z})
\label{EQq4.02}
\end{equation}
applied to the rows of the tensor generates continuity equations
\begin{equation}
\bar \partial_{\beta}T^{\alpha\beta}=0
\label{EQq4.03}
\end{equation}
for the electromagnetic energy and momentum.
The energy continuity equation 
and the momentum continuity equation
are mathematically equivalent to Poynting's theorem and the momentum 
continuity equation for a linear, isotropic, homogeneous, transparent,
dispersionless dielectric.
ii) Conservation of angular momentum requires diagonal symmetry
\begin{equation}
T^{\alpha\beta}=T^{\beta\alpha}.
\label{EQq4.04}
\end{equation}
iii) The array has a vanishing trace
\begin{equation}
T^{\alpha}_{\alpha}= 0
\label{EQq4.05}
\end{equation}
corresponding to massless particles \cite{BIJackson,BILL}.
iv) The components of the four-momentum
\begin{subequations}
\label{EQq4.06}
\begin{equation}
U=\int_{\sigma} dv T^{00}
\label{EQq4.06a}
\end{equation}
\begin{equation}
G^{i}=\frac{1}{c}\int_{\sigma} dv T^{0i}
\label{EQq4.06b}
\end{equation}
\end{subequations}
are conserved \cite{BILL} in an unimpeded flow.
\par

\includegraphics[scale=0.75]{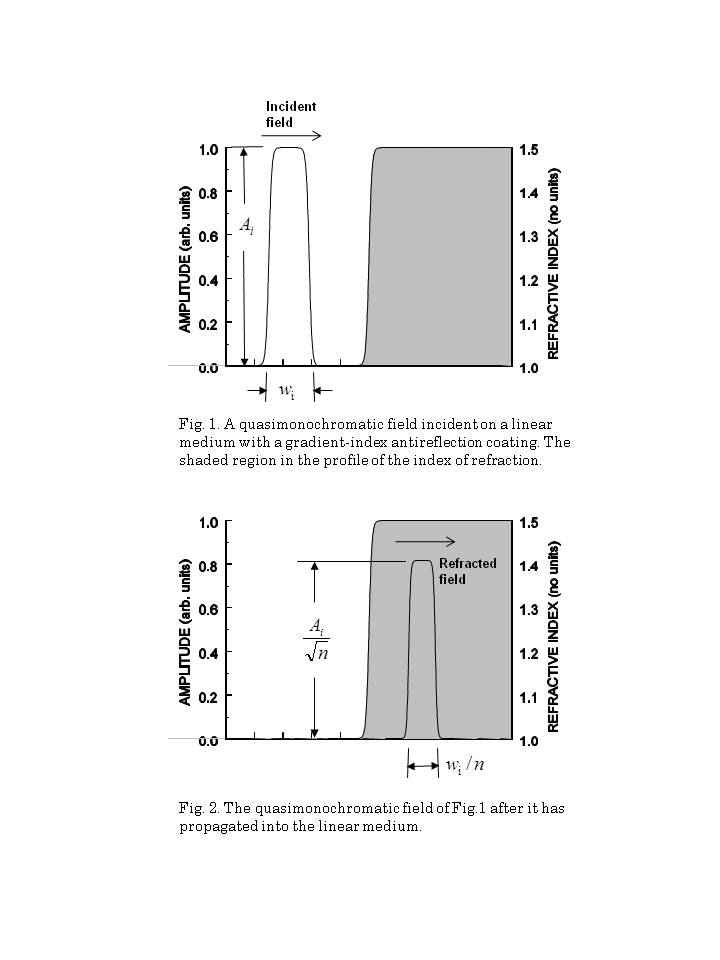}
\end{document}